\documentclass{emulateapj}

\begin{document}

\shortauthors{Moffat \& Toth} \shorttitle{Testing modified gravity with satellites}

\title{Testing modified gravity with motion of satellites around galaxies}

\author{J. W. Moffat$^{\dag *}$ and V. T. Toth$^{\dag}$\\~\\}

\affil{$^\dag$Perimeter Institute, 31 Caroline St North, Waterloo, Ontario N2L 2Y5, Canada}
\affil{$^*$Department of Physics, University of Waterloo, Waterloo, Ontario N2L 3G1, Canada}

\begin{abstract}
A modified gravity (MOG) theory, that has been successfully fitted to galaxy rotational velocity data, cluster data, the Bullet Cluster 1E0657-56 and cosmological observations, is shown to be in good agreement with the motion of satellite galaxies around host galaxies at distances 50-400~kpc.
\end{abstract}


\section{Introduction}
\label{sec:intro}

Dark matter has been used successfully to explain the rotational velocity data for galaxies and clusters of galaxies, lensing by galaxies and clusters of galaxies, and cosmological observations. Within the framework of general relativity (GR), this standard model of cosmology ($\Lambda$CDM) provides a successful description of the universe. The model predicts a universe that is composed of $\sim 4\%$ normal baryonic matter and $\sim 24\%$ non-baryonic cold dark matter (CDM), while the rest is made up of dark energy \citep{WMAP06}. However, dark matter has not been detected after much experimental effort. It is possible that exotic dark matter does not exist in sufficient quantities to explain the large amount of astrophysical and cosmological data. Instead, Newtonian gravity and GR must be modified to account for the data.

A relativistic MOG has been proposed to explain the astrophysical and large scale cosmological data \citep{Moffat2005,Moffat2006,Moffat2007,Moffat2007a,Moffat2007b} without exotic dark matter. The simplest version of the MOG is a scalar-tensor-vector gravity (STVG) theory \citep{Moffat2006,Moffat2007b}. The theory has been fitted remarkably well to the rotational velocity curve data of a large number of galaxies \citep{Brownstein2006a,Moffat2007b}, to the velocity dispersion of globular clusters \citep{Moffat2008} and to the mass and thermal profiles of clusters of galaxies \citep{Brownstein2006b,Moffat2007b}. It has also been fitted to the recent data of Clowe et al. \citep{Clowe2006,Markevitch2006} describing the dramatic merging of two clusters 1E0657-56 (Bullet Cluster) \citep{Brownstein2007}. A successful fit to the CMB acoustical power spectrum data and mass power spectrum data has been achieved with MOG and a possible explanation of the acceleration of the universe appears naturally in the gravity theory \citep{Moffat2007,Moffat2007a}.

Recently, \cite{Prada2007} presented an analysis of galaxy observations of the Sloan Digital Sky Survey (SDSS, {\tt http://www.sdss.org/}) to test gravity and dark matter in the peripheral parts of galaxies at distances 50-400~kpc from the centers of galaxies. This field of extragalactic astronomy provides one of the main arguments for the presence of dark matter \citep{ZW1994,Prada2003}. In the present work, we repeat part of this analysis and confirm that MOG predicts satellite galaxy velocities that are consistent with SDSS observations.

\section{Data Analysis}

\begin{figure*}[t]
\includegraphics[width=0.48\linewidth]{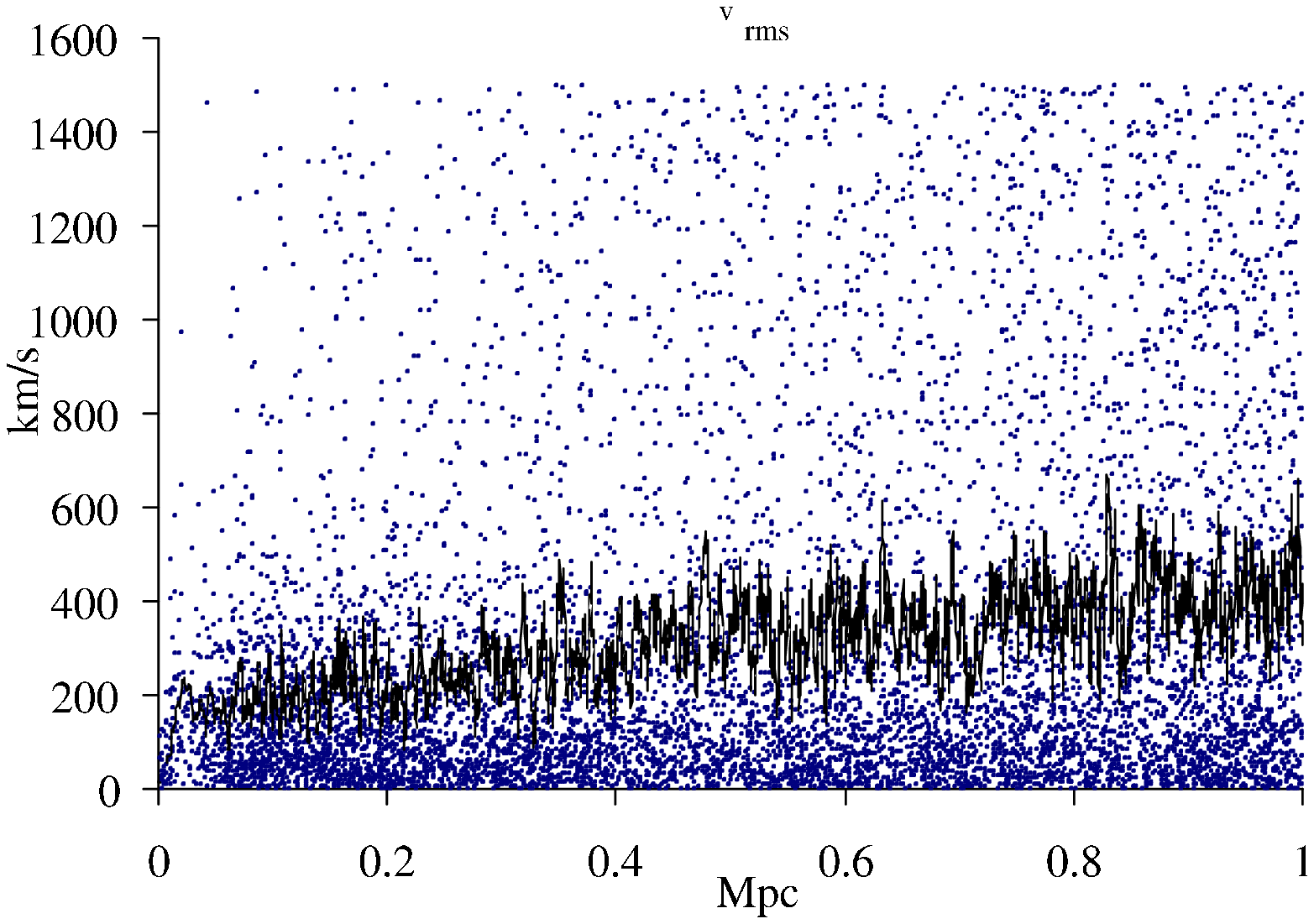}
\hskip 0.01\linewidth
\includegraphics[width=0.48\linewidth]{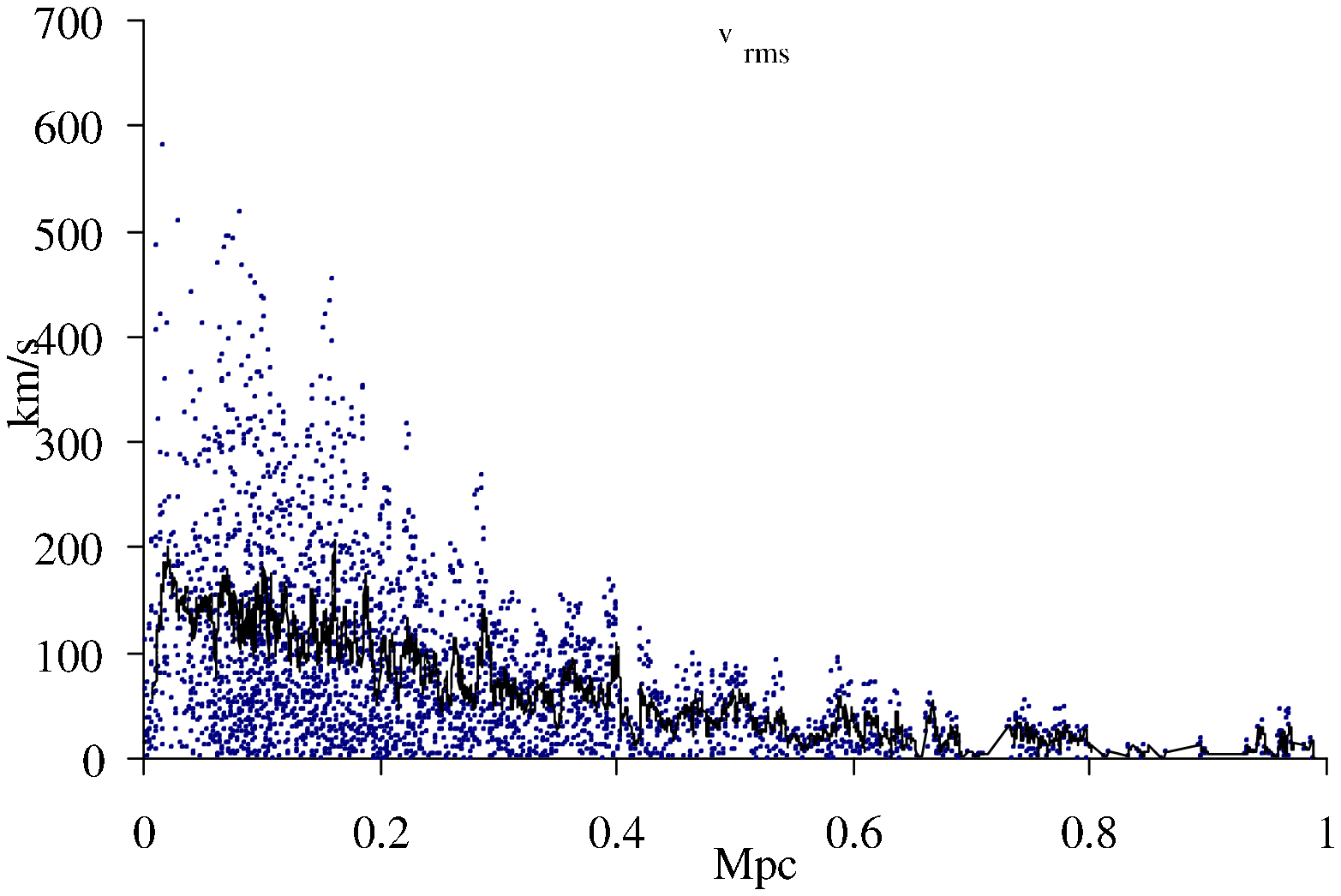}
\\
\includegraphics[width=0.48\linewidth]{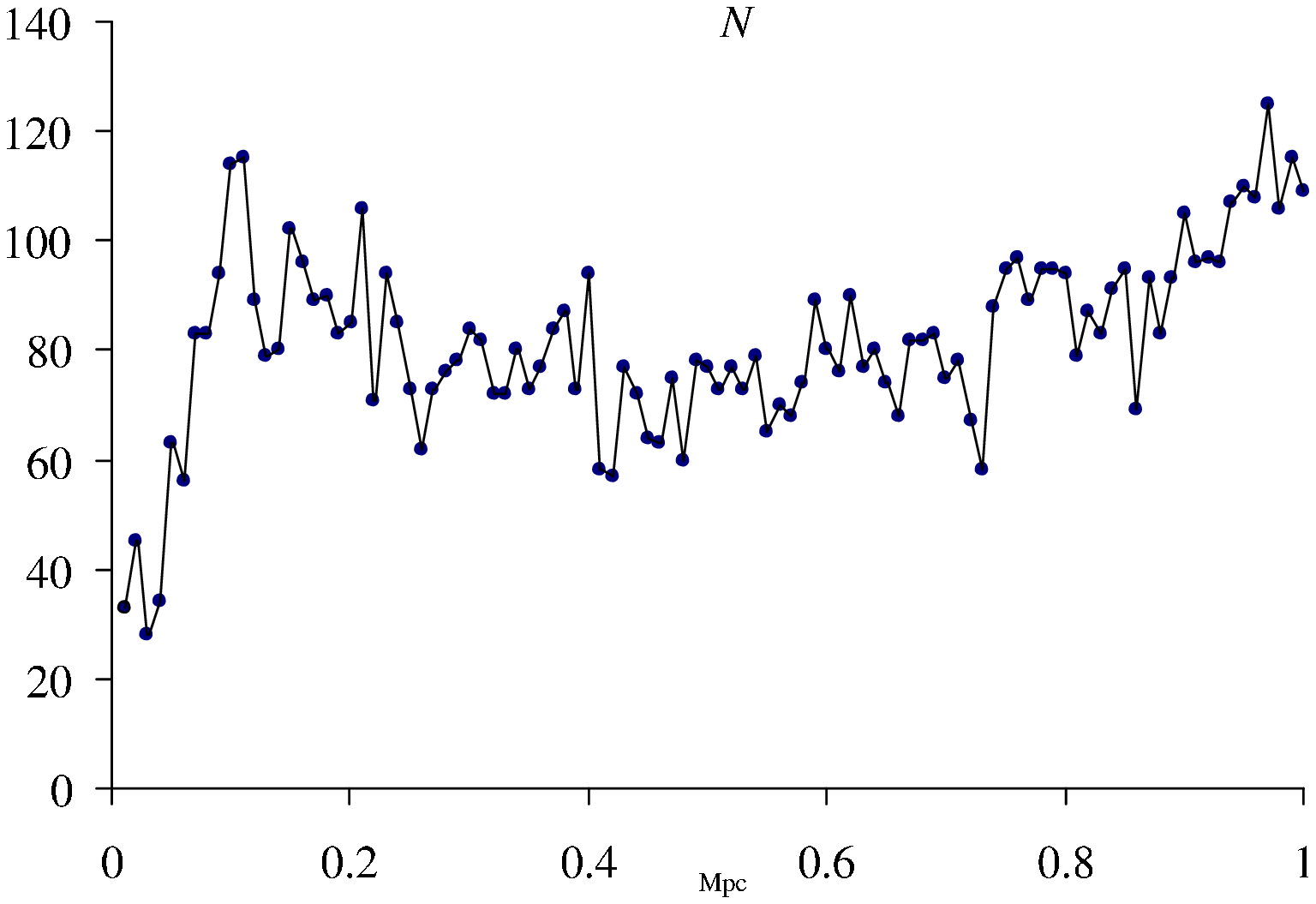}
\hskip 0.01\linewidth
\includegraphics[width=0.48\linewidth]{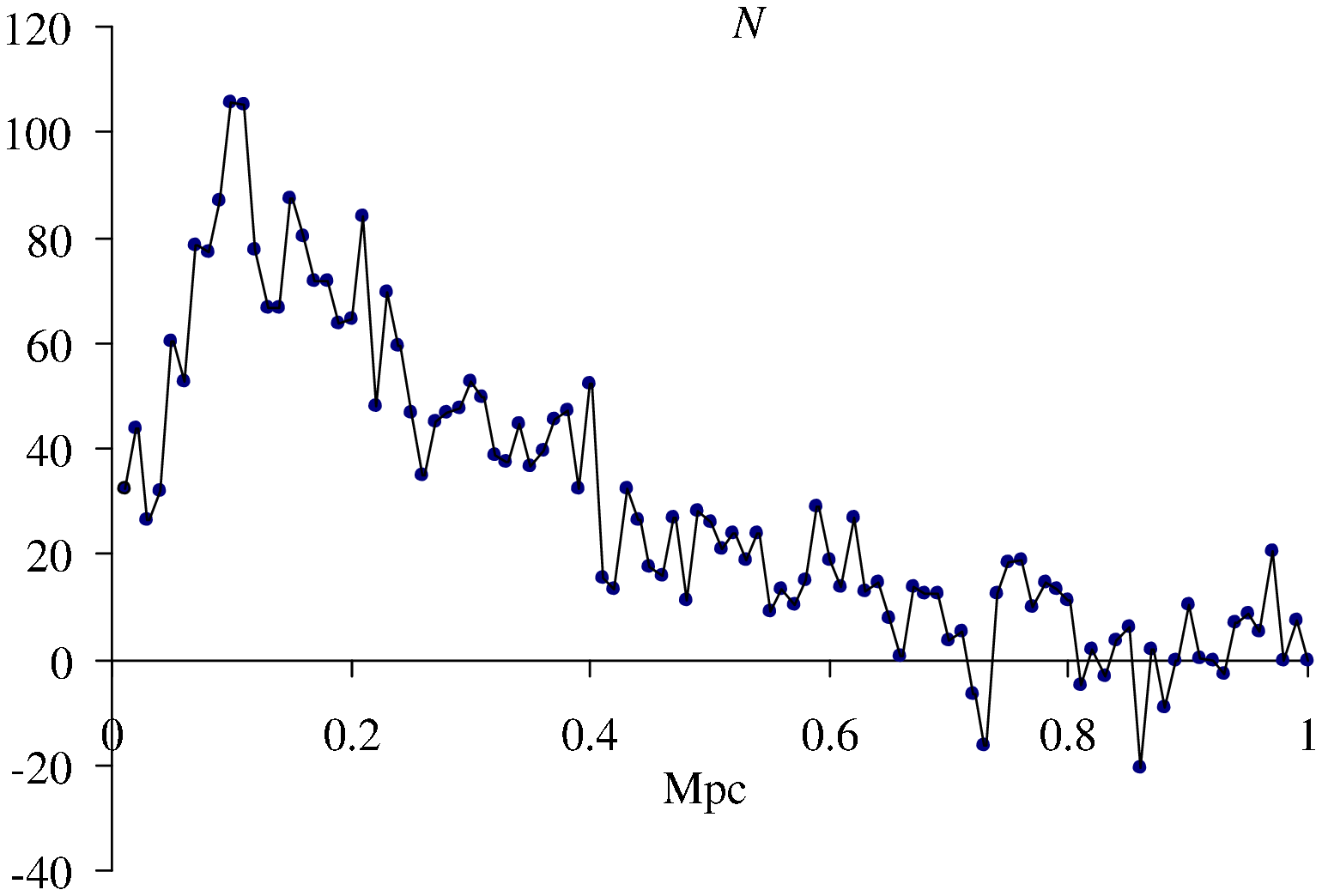}
\caption{Line-of-sight velocities (top left) and number densities (bottom left) of candidate satellite galaxies as a function of projected distance from the candidate host. After removal of candidate interlopers, the line-of-sight velocities (top right) and number densities (bottom right) are also shown. The solid lines in the top plots are moving window averages.}
\label{fig:plot1}
\end{figure*}

The SDSS provides imaging data over 9500~deg$^2$ in five photometric bands. Galaxy spectra are determined by the CCD imaging and the SDSS 2.5m telescope on Apache Point, New Mexico. Over half a million galaxies brighter than Petrosian $r$-magnitude $17.77$ over 7400~deg$^2$ are included in the SDSS data with a redshift accuracy better than 30~km/s.

We followed the analysis of \cite{Prada2007} and obtained a data set of nearly 700000 galaxies from the latest SDSS data release (Data Release 6, \cite{SDSSDR6}). We checked our query results by searching for, and locating several galaxies the parameters of which are known from the literature.

Rest frame absolute magnitudes can be computed in the SDSS $g$-band from the extinction-corrected apparent magnitudes assuming a Hubble constant $H_0\simeq 71~{\rm km s}^{-1} {\rm Mpc}^{-1}$. All magnitudes were $k$-corrected to $z=0$. SDSS redshifts are heliocentric; before calculating distances, we also took into account the solar system's motion relative to the CMB.

Due to the complications of calculating modified gravity for non-spherical objects, \cite{Prada2007} restricted their analysis only to red galaxies, the vast majority of which are either elliptical galaxies or are dominated by bulges. We followed a similar strategy, restricting our selection of candidate host galaxies to luminous red galaxies, as identified by \cite{SDSSDR6}. We also restricted our selection to galaxies with a recession velocity between 3000~km/s and 25000~km/s, which yielded approximately 234000 galaxies in total.

Still continuing with the strategy laid out by \cite{Prada2007}, we identified a host galaxy as a galaxy that is at least 4 times brighter than any other galaxy within a projected distance of $R=1$~Mpc and line-of-sight velocity within 1500~km/s that of the candidate host. Thereafter, we identified any galaxies at least 4 times fainter than the candidate host within a projected distance of 1~Mpc and with a line-of-sight velocity within 1500 km/s of that of the host as a candidate satellite galaxy. This strategy yielded a total of $\sim$3600 host galaxies with $\sim$8200 satellites. The procedure assigned a small percentage ($\sim 1.5\%$) of satellites to multiple hosts; we made no attempt to eliminate these from the data set.

The velocity distribution of candidate satellites as a function of projected distance from candidate hosts is shown in Figure~\ref{fig:plot1} (top left). We note that our choices of parameters (in particular, our choices of the factor four in the difference in luminosity between candidate hosts and satellites, or the restriction of hosts to luminous red galaxies) did not substantially affect the resulting distribution.

Obviously, the procedure outlined here does not guarantee that all candidate satellites are, in fact, true satellites of their respective candidate hosts. Indeed, one would expect a mix of satellite galaxies and a random background of interlopers. The number of interlopers should increase as a linear function of the projected distance, whereas the number of true satellites should peak at some typical satellite orbital radius. This expectation is confirmed if we plot the number density of candidate satellites as a function of projected distance (Figure~\ref{fig:plot1} bottom left). We are immediately presented with a possible editing strategy: after binning the data according to projected distance, we can remove a number of candidate satellites from the sample to account for the random background. By assuming that at 1~Mpc, none of the candidate satellites are, in fact, true satellites (i.e., all are random interlopers), we obtain a simple formulation: the number of galaxies to be removed from each bin is $N_{1\mathrm{Mpc}}R$, where $N_{1\mathrm{Mpc}}$ is the number of candidate satellites at 1~Mpc. The resulting number density of candidate satellites is shown in Figure~\ref{fig:plot1} (bottom right).

One question remains: what is a satellite? In other words, which galaxies are to be removed from the sample? We answer this question by noting that at a large projected distance from a candidate host, a candidate satellite is much less likely to be a true satellite of that host if its line-of-sight velocity also differs greatly from that of the host. Therefore, the candidate satellites that we remove from the sample are those with the greatest difference in line-of-sight velocity relative to the respective candidate host. The velocity distribution of the remaining satellites, shown in Figure~\ref{fig:plot1} (top right), is the data set that we study. As this simplistic selection procedure tends to underestimate satellite velocities at high projected distance, in the following, we shall restrict our investigation to a projected distance of 400~kpc or less.

\section{Modified Gravity Predictions}
\label{sec:results}

Predictions for modified gravity can be made by solving the Jeans equation, which gives the radial velocity dispersion $\sigma_r^2(r)$ as a function of radial distance. \citet{Prada2007} find that neither Newtonian gravity, nor MOND are compatible with observations. \citet{Angus2008}, however, demonstrated that a suitably chosen anisotropic model and appropriately chosen galaxy masses can be used to achieve a good fit for MOND.

Radial velocity dispersions in a spherically symmetric gravitational field can be computed using the Jeans equation \citep{Binney1987}:
\begin{equation}
\label{Jeans}
\frac{d(\nu\sigma_r^2)}{dr}+\frac{2\nu}{r}\beta\sigma_r^2=-\nu\frac{d\Phi}{dr},
\end{equation}
where $\nu$ is the spatial number density of particles, $v_r$ is the radial velocity, $\beta(r)=1-[\sigma_\theta^2(r)+\sigma_\phi^2(r)]/2\sigma_r^2(r)$ is the velocity anisotropy, $\Phi(r)$ is the gravitational potential, and we are using spherical coordinates $r$, $\theta$, $\phi$. We can write Eq.(\ref{Jeans}) in the form
\begin{equation}
\label{Jeans2} \frac{d\sigma_r^2}{dr}+\frac{A\sigma_r^2}{r}=-g(r).
\end{equation}
Here, we have
\begin{equation}
A=2\beta(r)+\gamma(r),
\end{equation}
where $\gamma(r)=d\ln\nu(r)/d\ln{r}$.

If we assume that the velocity distribution of satellite galaxies is isotropic, $\beta=0$. In general, $\beta$ needs to be neither zero nor constant. The number density of candidate satellites favors a value of $\gamma\simeq -2.5$. We note that our MOG predictions are not particularly sensitive to these parameters; no ``fine-tuning'' is needed to obtain good agreement with the data.

The observed velocity dispersion is along the observer's line of sight, seen as a function of the projected distance from the host galaxy. Therefore, it is necessary to integrate velocities along the line-of-sight:
\begin{equation}
\sigma_\mathrm{LOS}^2(R)=\frac{\int\limits_0^\infty\left[y^2+(1-\beta)R^2\right]r^{-2}\sigma_r^2(y)\nu(y)~dy}
{\int\limits_0^\infty\nu(y)~dy},
\end{equation}
where $\nu$ is the spatial number density of satellite galaxies as a function of distance from the host galaxy, and $y$ is related to the projected distance $R$ and 3-dimensional distance $r$ by
\begin{equation}
r^2=R^2+y^2.
\end{equation}
Changing integration variables to eliminate $y$, we can express the observed line-of-sight velocity dispersion as a function of projected distance as
\begin{equation}
\sigma_\mathrm{LOS}^2(R)=\frac{\int\limits_R^\infty(r^2-\beta R^2)\sigma_r^2(r)\nu(r)/r\sqrt{r^2-R^2}~dr}
{\int\limits_R^\infty r\nu(r)/\sqrt{r^2-R^2}~dr}.
\label{eq:LOS}
\end{equation}

From the field equations derived from the STVG action, we obtain the modified Newtonian acceleration law for weak gravitational fields~\citep{Moffat2006,Moffat2007b} of a point source with mass $M$:
\begin{equation}
\label{MOGpot} g_{\rm MOG}(r)=\frac{G_NM}{r^2}\left\{1+\alpha\left[1-e^{-\mu r}\left(1+\mu r\right)\right]\right\},
\end{equation}
where $G_N$ is the Newtonian gravitational constant, while the MOG parameters $\alpha$ and $\mu$ determine the coupling strength of the ``fifth force'' vector $\phi_\mu$ to baryon matter and the range of the force, respectively.

\begin{figure}[t]
\includegraphics[width=\linewidth]{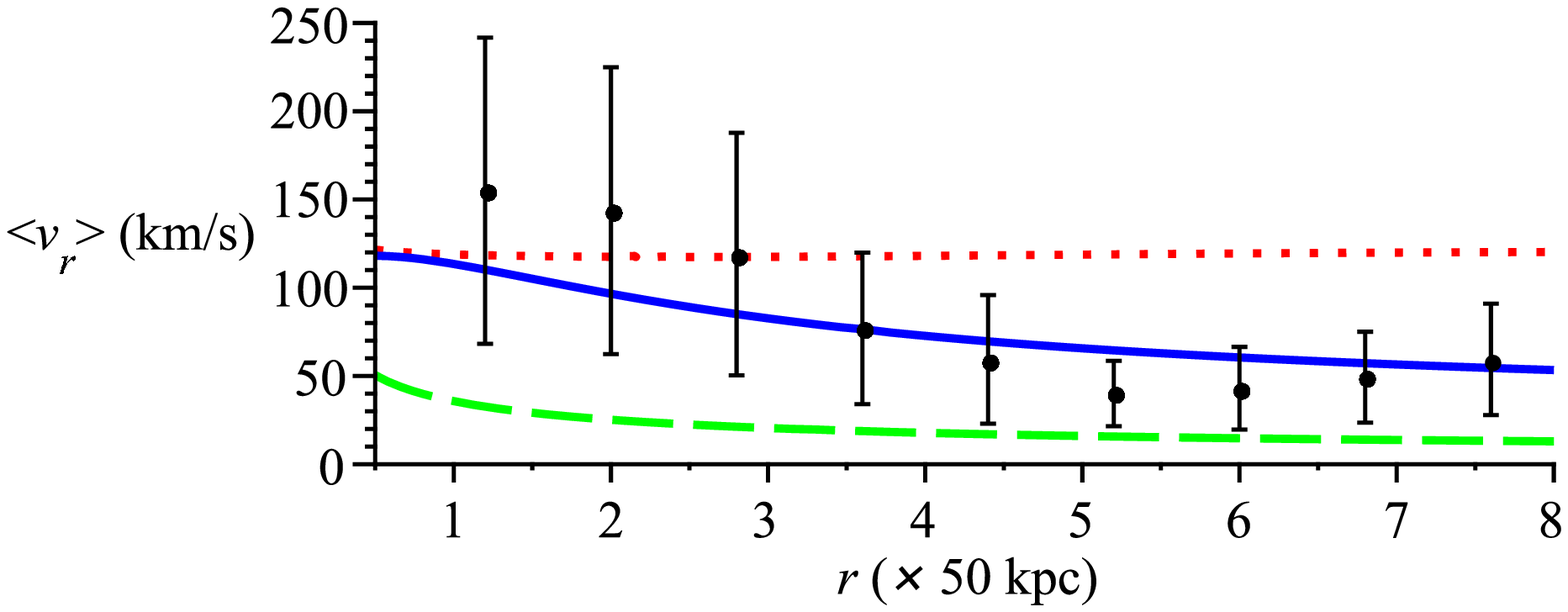}
\includegraphics[width=\linewidth]{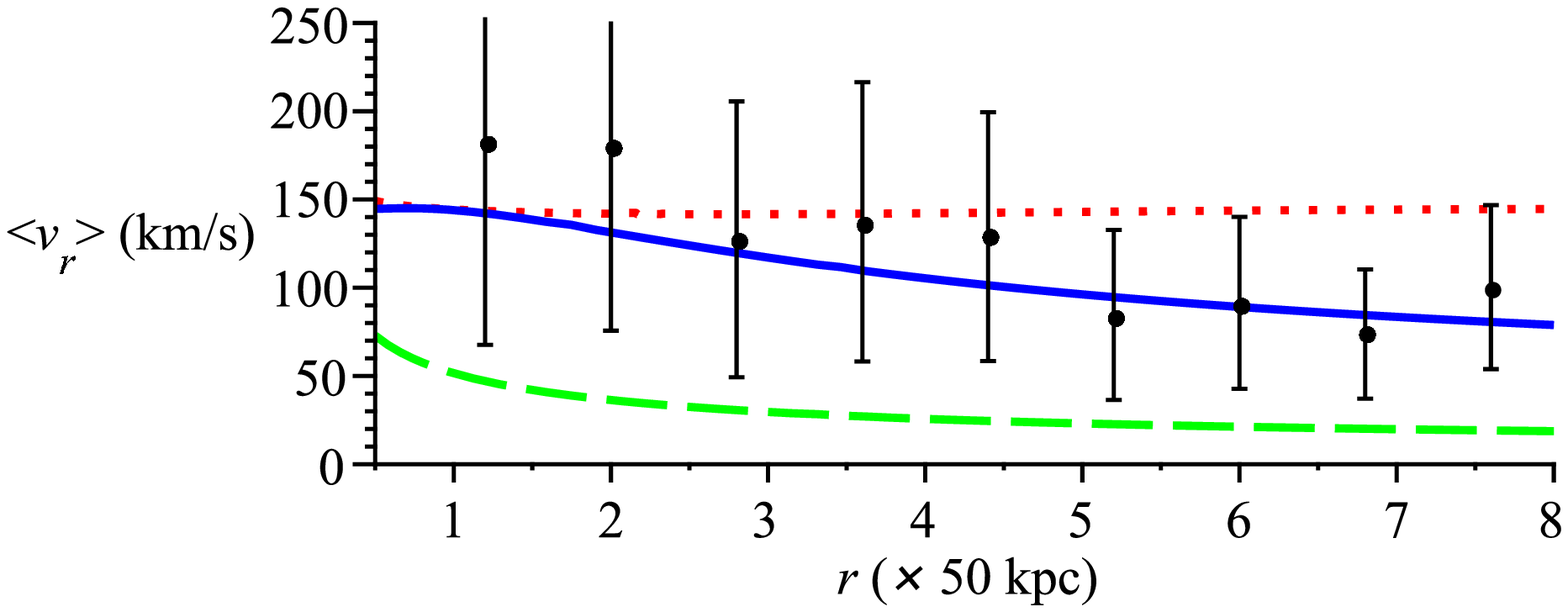}
\caption{Root-mean-square line-of-sight velocities and theoretical predictions. Line of sight velocities are binned by projected distance, in bins of 40~kpc. The parameter-free MOG prediction (solid blue line) using nominal host galaxy masses is in good agreement with the data. MOND (dotted red line) is in marginal agreement, which can be improved by, for instance, varying the parameters $\beta$ and $\gamma$ with radial distance. The same is not true for the Newtonian prediction (dashed green line). The two panels show two host galaxy luminosity ranges. Top: $-20.5>M^1_g>-21.1$, bottom: $-21.1>M^2_g>-21.6$.}
\label{fig:plot2}
\end{figure}

In recent work \citep{Moffat2007b}, we have been able to develop formulae that predict the values of the $\alpha$ and $\mu$ parameters, in the form
\begin{eqnarray}
\mu&\simeq&\frac{D}{\sqrt{M}},\label{eq:mu}\\
\alpha&\simeq&\frac{M}{(\sqrt{M}+E)^2}\left(\frac{G_\infty}{G_N}-1\right),\label{eq:alpha}
\end{eqnarray}
where the parameters
\begin{eqnarray}
G_\infty&\simeq&20G_N,\\
D&\simeq&6250~M_\odot^{1/2}\mathrm{kpc}^{-1},\\
E&\simeq&25000~M_\odot^{1/2}
\end{eqnarray}
are determined from galaxy rotation curves and cosmological observations \citep{Moffat2007b}.

The MOND acceleration $g_{\rm MOND}$ is given by the solution of the non-linear equation
\begin{equation}
 g_{\rm MOND}\mu\left(\frac{|g_{\rm MOND}|}{a_0}\right)=\frac{G_NM(r)}{r^2},
\end{equation}
where $M$ is the mass of only baryons and $a_0=1.2\times 10^{-8}{\rm cm}\: {\rm sec}^{-2}$. The form of the function $\mu(x)$ originally proposed by Milgrom~\citep{Milgrom} is given by $\mu(x)=x/\sqrt{1+x^2}$; however, better fits and better asymptotic behavior are achieved using $\mu(x)=x/(1+x)$~\citep{Prada2007}.

Following in the footsteps of \cite{Prada2007}, we grouped satellite galaxy velocities for host galaxies in two luminosity ranges: $-20.5>M^1_g>-21.1$, and $-21.1>M^2_g>-21.6$. The corresponding masses for the host galaxies, calculated by \cite{Prada2007} on the basis of the work of \cite{Bell2001}, are
\begin{eqnarray}
M^1_*=7.2\times 10^{10}~M_\odot&&(-20.5>M^1_g>-21.1),\\
M^2_*=1.5\times 10^{11}~M_\odot&&(-21.1>M^2_g>-21.6).
\end{eqnarray}

We plotted the MOG prediction using these nominal masses and the values of $\beta=0$, $\gamma=-2.5$, along with the predictions of MOND and Newtonian gravity using the same parameters. As Figure~\ref{fig:plot2} shows, the MOG prediction is consistent with observational data, although MOG prefers a somewhat higher than nominal mass for the hosts; a least-squares fit yields $M^1_*\simeq 9.3\times~10^{10}~M_\odot$ and $M^2_*\simeq 2.2\times~10^{11}~M_\odot$.

\section{Conclusions}
\label{sec:conclusions}

Observational data presented by \cite{SDSSDR6} and studied by \cite{Prada2007} are viewed as evidence of the success of the $\Lambda$CDM model. However, the data are also in excellent agreement with the predictions of the parameter-free solutions of our modified gravity theory, without postulating a peculiarly distributed halo of exotic dark matter.

\section*{Acknowledgments}

The research was partially supported by National Research Council of Canada. Research at the Perimeter Institute for Theoretical Physics is supported by the Government of Canada through NSERC and by the Province of Ontario through the Ministry of Research and Innovation (MRI).

\bibliographystyle{apj}

\end{document}